\documentclass[twocolumn,showpacs,preprintnumbers,amsmath,amssymb,superscriptaddress]{revtex4}

\usepackage{graphicx}
\usepackage{dcolumn}
\usepackage{subfigure}
\usepackage{bm}

\newcommand{\omqls}{\Omega_Q^{ls}}
\newcommand{\mmc}{MCMC}

\begin{document}

\title{Cosmic Microwave Background and Supernova 
Constraints on Quintessence: Concordance
Regions and Target Models}

\author{Robert R. Caldwell}
\affiliation{Department of Physics \& Astronomy, Dartmouth College,
6127 Wilder Laboratory,
Hanover, NH 03755}

\author{Michael Doran}
\affiliation{Department of Physics \& Astronomy, Dartmouth College,
6127 Wilder Laboratory,
Hanover, NH 03755}

\date{\today} 


\begin{abstract}
 
We perform a detailed comparison of the Wilkinson Microwave Anisotropy Probe
(WMAP) measurements of the cosmic microwave background (CMB) temperature and
polarization anisotropy with the predictions of quintessence cosmological
models of dark energy. We consider a wide range of quintessence models,
including: a constant equation-of-state;  a simply-parametrized, time-evolving
equation-of-state;  a class of models of early quintessence; scalar fields with
an inverse-power law potential. We also provide a joint fit to the CBI and
ACBAR CMB data, and the type 1a supernovae. Using these select constraints we
identify viable, target models for further analysis.

\end{abstract}

\maketitle



The precision measurement of the cosmic microwave background (CMB) by the
Wilkinson Microwave Anisotropy Probe (WMAP) satellite
\cite{MAPmission,MAPresults} represents a milestone in experimental cosmology.
Designed for precision measurement of the CMB anisotropy on angular scales from
the full sky down to several arc minutes, this ongoing mission has already
provided a sharp record of the conditions in the Universe from the epoch of
last scattering to the present. In light of this powerful data
\cite{Hinshaw:2003ex,Kogut:2003et,Page:2003fa,Spergel:2003cb,Verde:2003ey}, we
must consider anew our cosmological theories.

\begin{figure}[t]
\scalebox{0.3}{\includegraphics{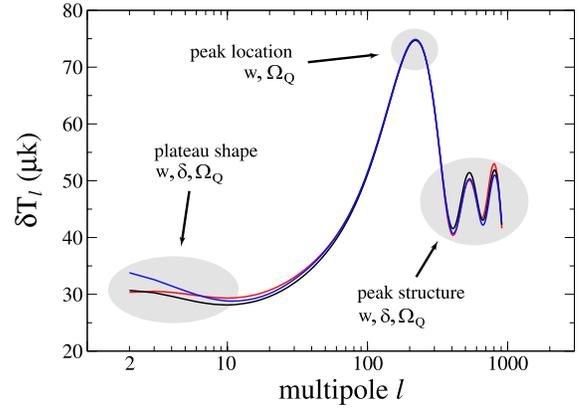}}
\caption{\label{fig:cmb_bw} 
The pattern of CMB anisotropy can reveal information about the quintessence
abundance ($\Omega_Q$), equation-of-state ($w$), and behavior of fluctuations
($\delta$). The three curves are examples of constant equation-of-state models
which differ little by eye, but are distinguished by the data. The red
($w=-0.5$) and blue ($-1.2$) curves are both low-$\chi^2$
CMB-indistinguishable, but distinct with respect to SNe. The black curve
($-0.8$), although it is consistent with the SNe data and matches the  
location and height of the first acoustic peak determined by WMAP
\cite{Page:2003fa}, is rejected by the CMB at the $3\sigma$-level.} 
\end{figure}

We aim to use the WMAP results to test cosmological theories of the
accelerating Universe --- to seek clues to the nature of the dark energy.
Despite the absence of a direct dark-energy interaction with our baryonic
world, the CMB photons provide a probe of the presence of the dark energy,
complementary to the type 1a supernovae. Via the integrated Sachs-Wolfe effect
on large angular scales, the geometric optics of the last-scattering sound
horizon on degree scales, and the pattern of acoustic oscillations on smaller
angular scales, we expect the CMB to reveal information about the dark energy
density, equation-of-state, and behavior of fluctuations. These effects are
illustrated in Figure~\ref{fig:cmb_bw}.

In this article we test the {\it quintessence hypothesis}, that a dynamical,
time-evolving, negative pressure, inhomogeneous form of energy dominates the
cosmic energy density and is responsible for the cosmic acceleration
\cite{Ratra:1987rm,Peebles:1987ek,Wetterich:fm,Wetterich:bg,Coble:1996te,Caldwell:1997ii,Turner:1998ex}.
To be precise, we carry out an extensive analysis of the CMB anisotropy and
mass fluctuation spectra for a wide range of quintessence models. These models
are: (Q1) models with a constant equation-of-state, $w$, including $w<-1$; (Q2)
models with a simply-parametrized, time-evolving $w$; (Q3) early quintessence
models, with a non-negligible energy density during the recombination era; (Q4)
trackers described by a scalar field evolving under an inverse-power law
potential.  


The suite of parameters describing the cosmological models are split into
spacetime plus ``matter sector" variables, $\theta_{M}$, and separate
quintessence parameters, $\theta_{Q}$. The spacetime and ``matter sector" of
the quintessence models are specified by the parameter set $\theta_{M} =
\{\Omega_b h^2,\, \Omega_{cdm} h^2,\, h,\, n_s,\, A_S,\, \tau_{r}\}$. In order,
these are the baryon density, cold dark matter density, hubble parameter,
scalar perturbation spectral index, scalar perturbation amplitude, and optical
depth. In this investigation we restrict our attention to spatially-flat, cold
dark matter models with a primordial spectrum of nearly scale-invariant density
perturbations generated by inflation.  

The quintessence parameters vary from model to model. For the simplest family
of models, with a constant equation-of-state, we need only to specify
$\theta_{Q} = \{w\}$. For models which feature more realistic time-evolution of
the quintessence, which may include a non-negligible fraction of quintessence
at early times, more parameters are required, {\it e.g.} $\theta_{Q} = \{w,\,
dw/da,\,...\}$, to characterize the impact on the cosmology in general and the
CMB in particular.


Our analysis method is as follows: (i) compute the CMB and fluctuation power
spectra for a given cosmological model; (ii) compute the relative likelihood of
the model with respect to the experimental data; (iii) assemble the likelihood
function in parameter space to determine the range of viable quintessence
models. For step (i) we use both a version of CMBfast \cite{CMBfast}  modified
for quintessence, as well as the newly-available CMBEASY \cite{CMBEASY}. For
step (ii) we supplement the WMAP data with the complementary ACBAR \cite{ACBAR}
and CBI-MOSAIC data \cite{CBI_mosaic,CBI_deep} (using the same bins as Ref.
\cite{Spergel:2003cb,Verde:2003ey}), in addition to the current type 1a SNe
data \cite{Schmidt:1998ys,Riess:1998cb,Garnavich:1998th,Perlmutter:1998np}.
Certain other constraints, such as the HST Key Project measurement of $H_0$
\cite{Freedman:2000cf} or the limit from Big Bang Nucleosynthesis on  $\Omega_b
h^2$ \cite{Burles:2000zk} through the deuterium abundance measurement are
satisfied as cross-checks. It is remarkable that such agreement can be found
between such diverse phenomena. Our focus in the following investigation,
however, is primarily on the CMB and SNe.


\begin{figure}[h]
\scalebox{0.4}{\includegraphics{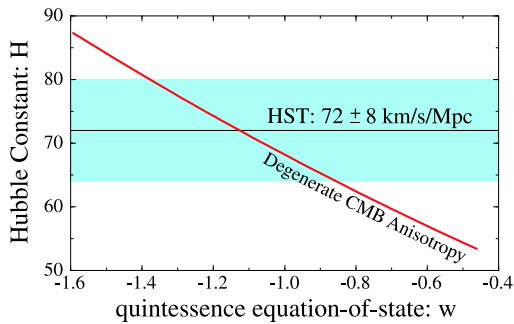}}
\caption{\label{fig:HubbleFit} 
The one-parameter family of best-fit models, which exploit the geometric
degeneracy of the CMB anisotropy pattern, is shown as the thick, red curve in
the $w-h$ plane. We have explored models in a six-dimensional cylinder in the
parameter space surrounding this ``best-fit line.''  The HST-Key Project
$1\sigma$ measurement of the Hubble constant is shown by the shaded band.} 
\end{figure}

\begin{figure}[h]
\scalebox{0.4}{\includegraphics{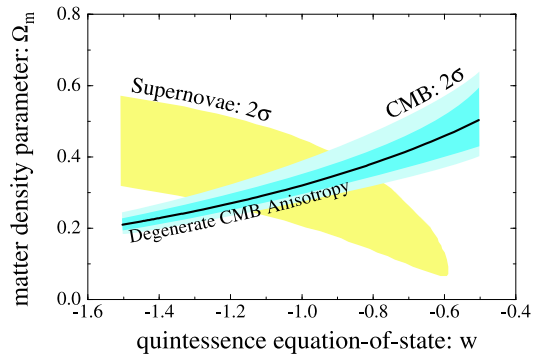}}
\caption{\label{fig:BestFit} 
The constraints on constant equation-of-state models due to CMB (WMAP, ACBAR,
CBI) and type 1a supernovae (Hi-Z, SCP) are shown. The starting point for our
parameter-search, the family of CMB-degenerate models, is shown by the thick,
black line.} 
\end{figure}

\noindent{\it Q1:} We have analyzed the cosmological constraints on the
simplest model of quintessence, characterized by a constant equation-of-state,
$w$. We have used the equivalence between a scalar field $\varphi$ with
potential $V(\varphi)$ and the equation-of-state $w$ in order to
self-consistently evaluate the quintessence fluctuations. For the range $w<-1$
we employ a {\it k}~essence model, keeping the sound speed (actually, this is
$d\omega^2/dk^2$) fixed at $c_s^2 = 1$. Since this model introduces only one
additional parameter beyond the basic set of spacetime plus matter sector
variables, we adopt a simplistic grid-based search for viable models. The
acceptance criteria for the Q1 models is based on a $\Delta\chi^2$-test. The
results of our survey of Q1 models are shown in
Figures~\ref{fig:HubbleFit},\,\ref{fig:BestFit}. We have exploited the
degeneracy of the CMB anisotropy pattern among models with the same apparent
angular size of the last scattering horizon  \cite{Huey:1998se}. Hence, there
is a family of models with $\Omega_b h^2 = 0.023$, $\Omega_{cdm} h^2 = 0.126$,
$n_s = 0.97$, and characterized by pairs $\{w,\ h\}$ having (nearly)
indistinguishable CMB anisotropy patterns. The pairs $\{w,\ h\}$ are shown in
Figure~\ref{fig:HubbleFit}, and all represent quintessence models with
$\chi^2=1429$ for the WMAP temperature-temperature and temperature-polarization
data --- a one-parameter family of best-fit models. From this starting point,
we explored over $6\times 10^4$ models distributed on a grid filling a
6-cylinder around the best-fit line, varying $\{\Omega_b h^2,\, \Omega_{cdm}
h^2,\, h,\, n_s,\, \tau_{r}\}$ at intervals in $w$. For each model we evaluate
the likelihood relative to WMAP, as well as the complementary ACBAR and
CBI-MOSAIC data. The $2\sigma$ boundary, based on a $\Delta\chi^2$ test for six
degrees of freedom is shown in Figure~\ref{fig:BestFit}. We have also evaluated
the constraint in the $w-\Omega_m$ plane for the combined Hi-Z/SCP type 1a
supernova data set, showing the $2\sigma$ region based on a $\Delta\chi^2$ test
for two degrees of freedom. Our basic conclusion from the overlapping
constraint regions is that there exist concordant models with $-1.25 \lesssim w
\lesssim -0.8$ and $0.25 \lesssim \Omega_{m} \lesssim 0.4$. We have identified
four sample models in Table~\ref{q1models} for further analysis.

\begin{table}[h] 
\begin{ruledtabular}
  \begin{tabular}{cccc}
   model & $w$  & $h$ & $\sigma_8$ \\ \hline
   Q1.1 & -0.82 & 0.630 & 0.84 \\
   Q1.2 & -1.00 & 0.682 & 0.89 \\
   Q1.3 & -1.18 & 0.737 & 0.96 \\
   Q1.4 & -1.25 & 0.759 & 0.97 \\
  \end{tabular}
\end{ruledtabular}
\caption{\label{q1models}Sample best-fit models with a constant 
equation-of-state (Q1). All models have $\Omega_b h^2 = 0.023$, $\Omega_{cdm} h^2 =
0.126$, $n_s=0.97$, and $\tau_{r} =0.11$.}
\end{table}


\newcommand{\threescale}{0.24}
\begin{figure*}[t]
\subfigure[Q2: monotonic evolution $w(a)$]{\label{fig::monotonic1}
  \includegraphics[scale =\threescale,angle=-90]{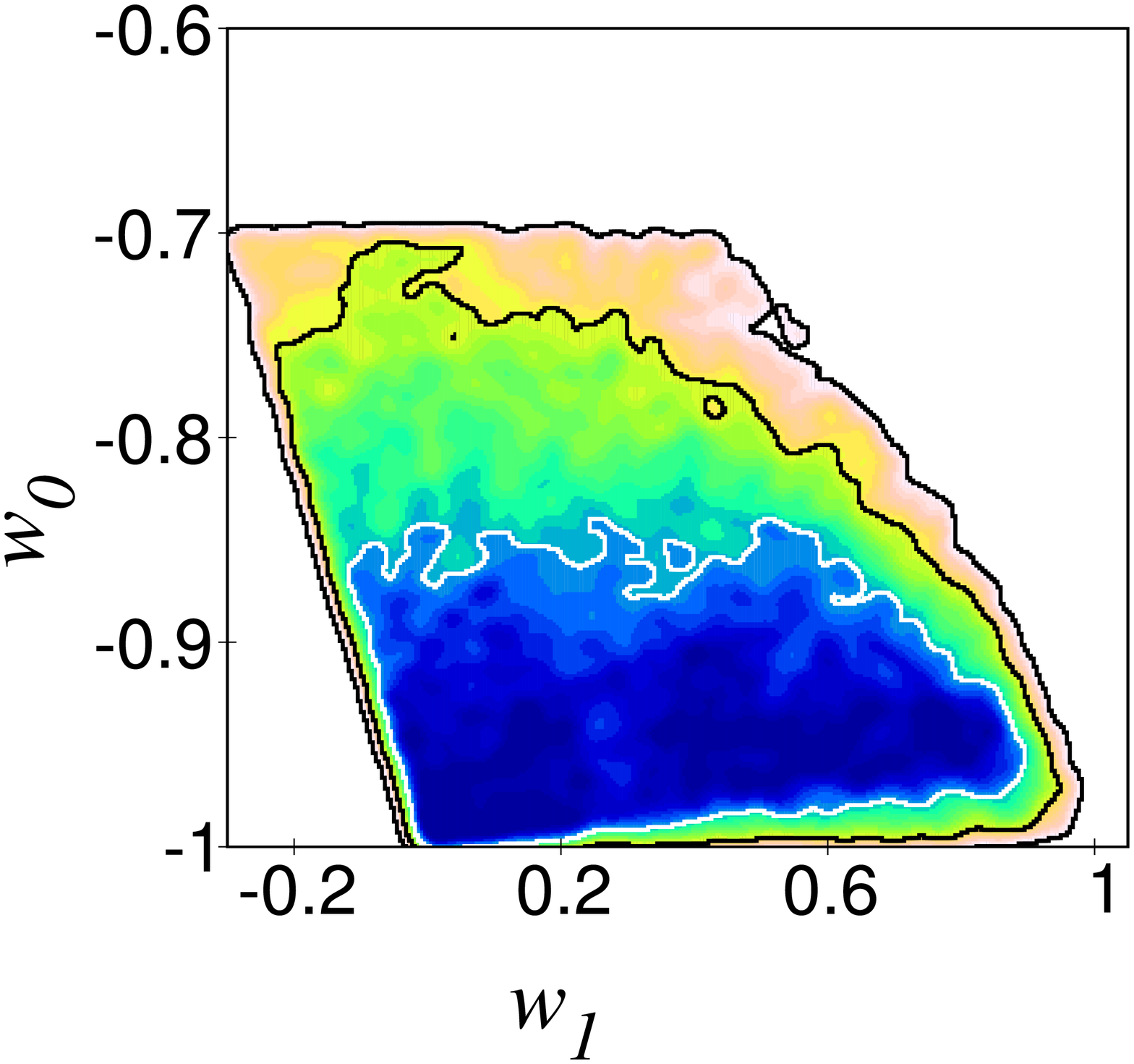}
}
\subfigure[Q3: leaping kinetic quintessence]{ \label{fig::leaping1}
  \includegraphics[scale =\threescale,angle=-90]{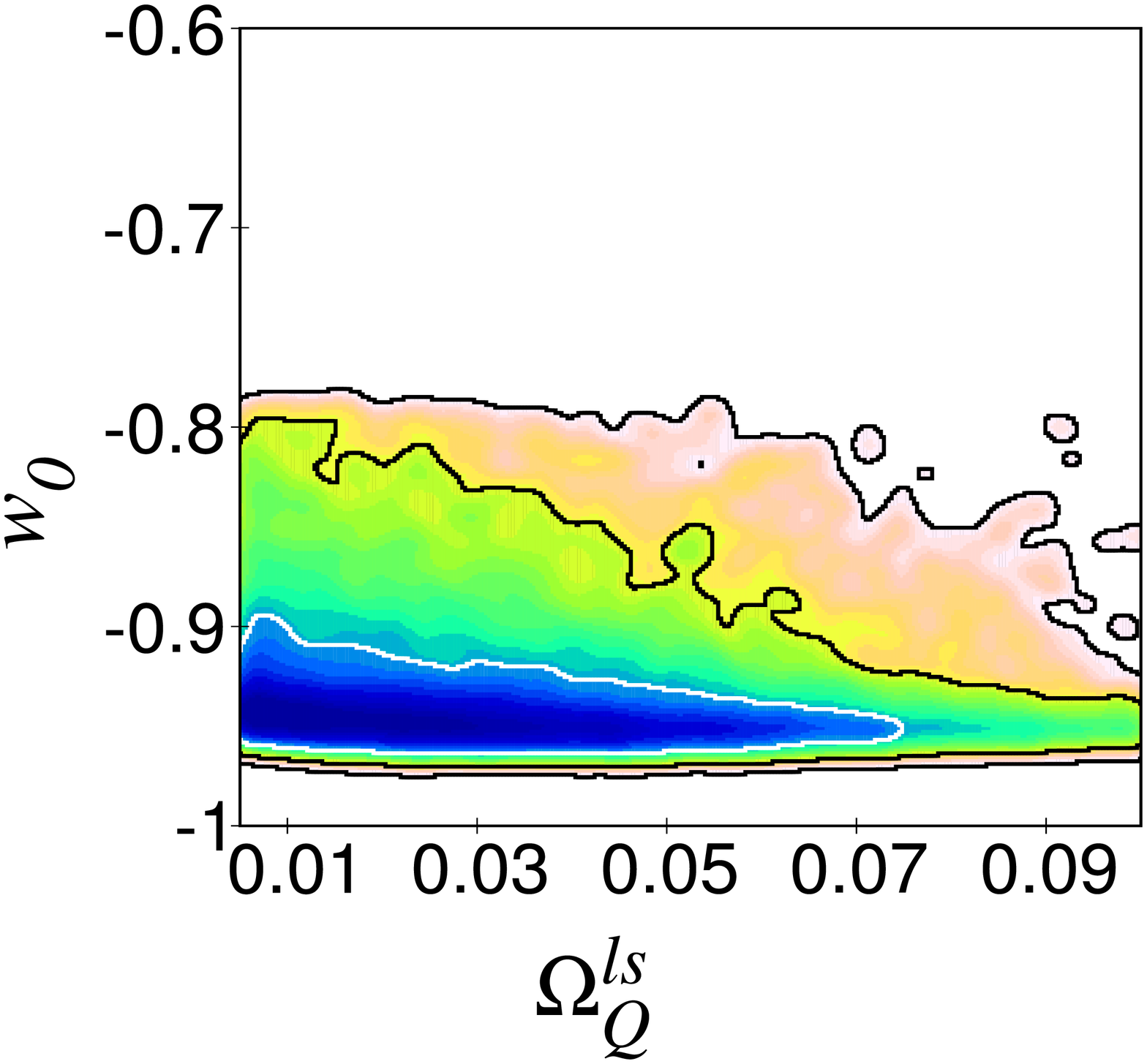}
}
\subfigure[Q4: inverse-power law]{ \label{fig::ipl1}
  \includegraphics[scale =\threescale,angle=-90]{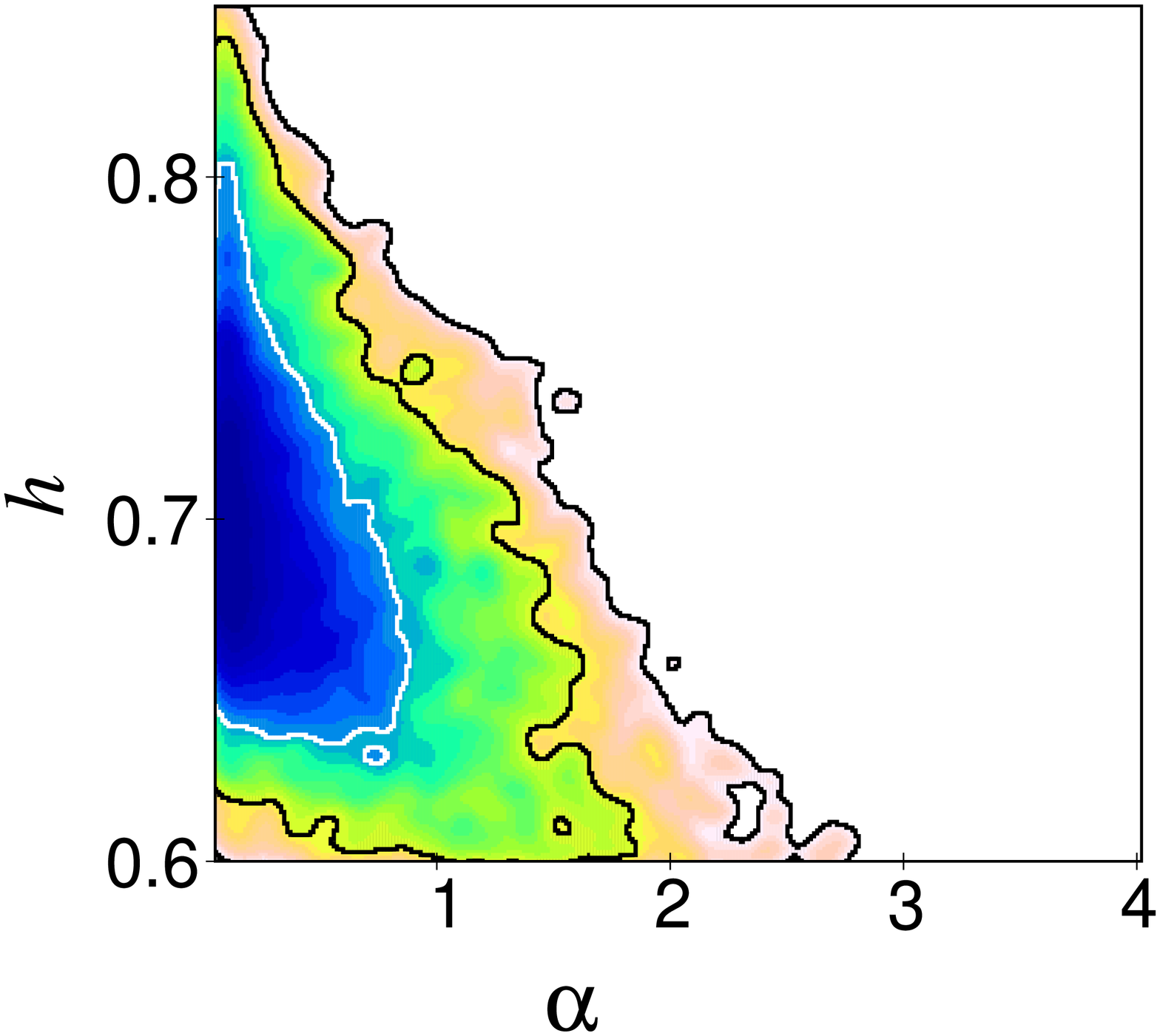}
}

\subfigure[Q2: monotonic evolution $w(a)$]{ \label{fig::monotonic2}
  \includegraphics[scale =\threescale,angle=-90]{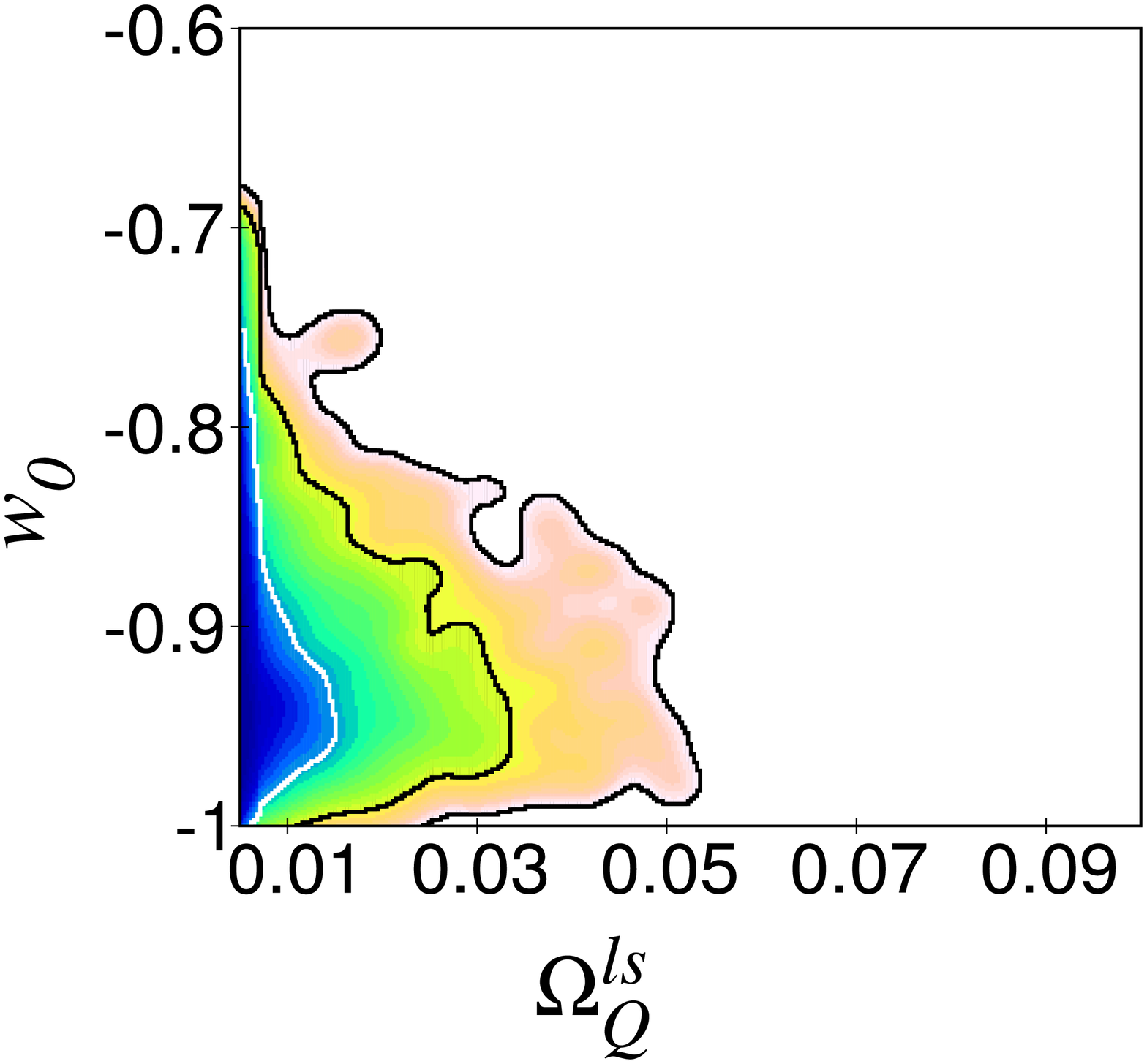}
}
\subfigure[Q3: leaping kinetic quintessence]{ \label{fig::leaping2}
  \includegraphics[scale =\threescale,angle=-90]{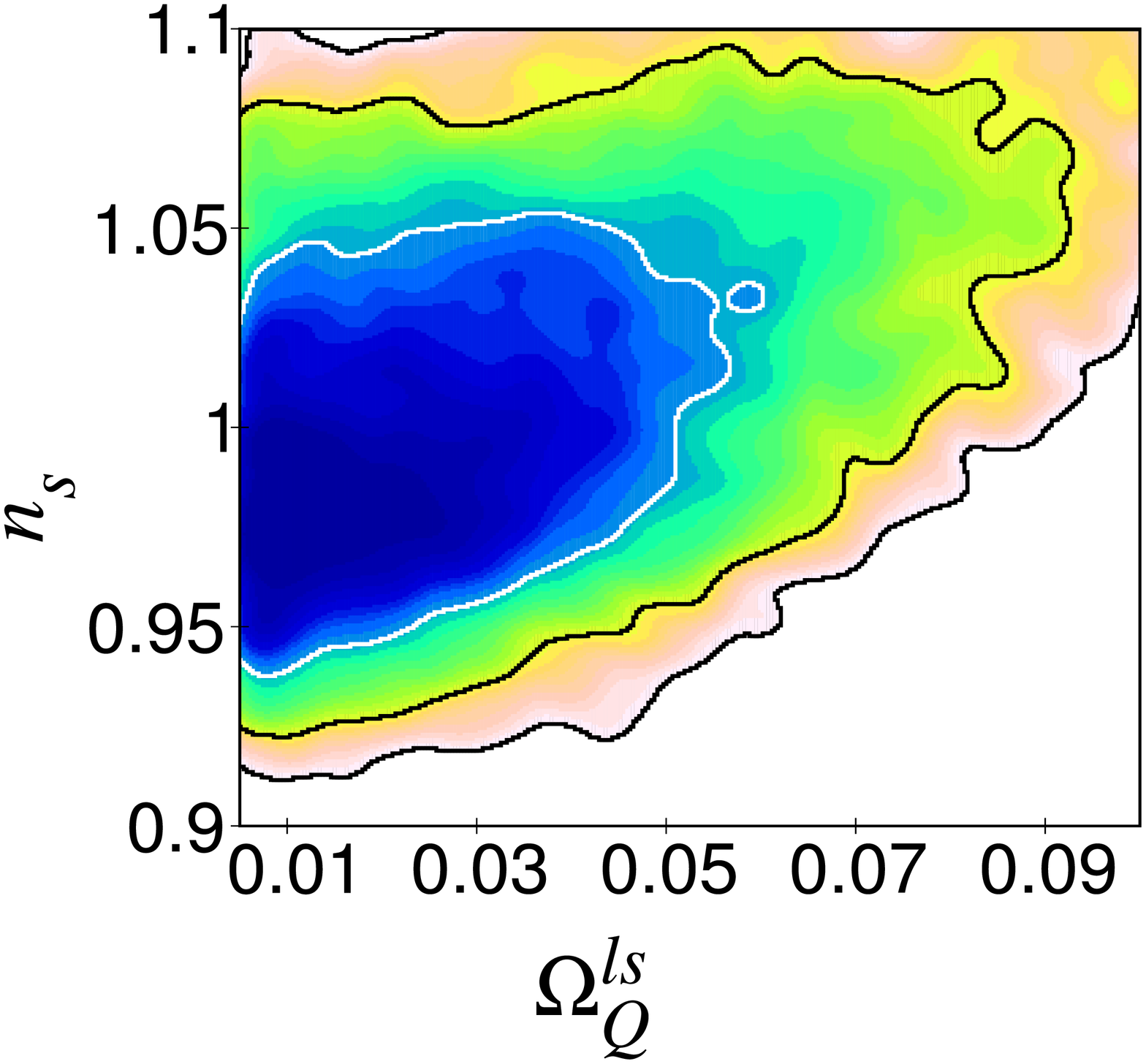}
}
\subfigure[Q4: inverse-power law]{\label{fig::ipl2}
  \includegraphics[scale =\threescale,angle=-90]{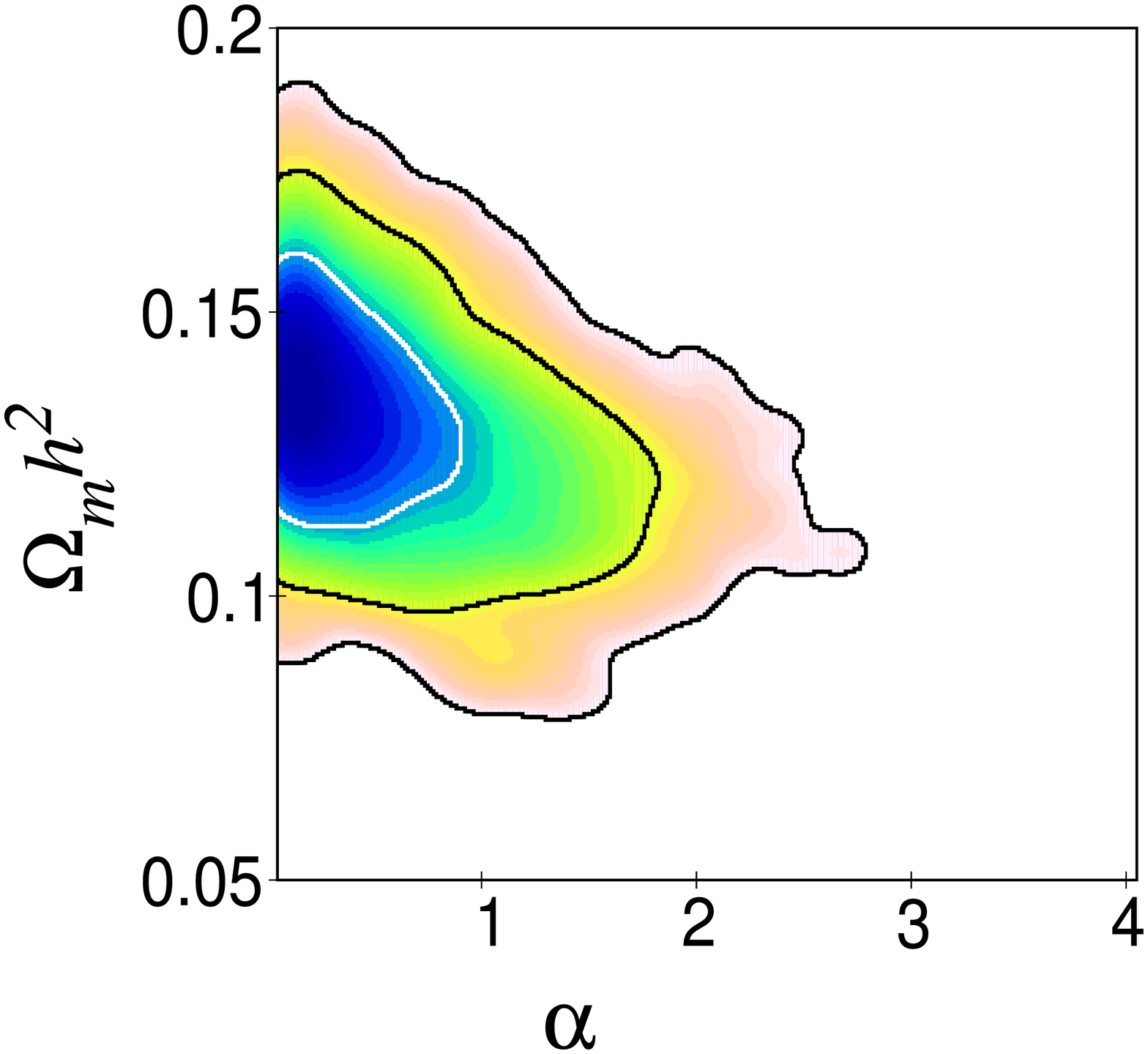}
}
\caption{The results of our \mmc\ search of the multi-dimensional parameter
search, for models Q2-4, are illustrated in the figures above. In all cases, we
have marginalized over the suppressed parameters. The solid lines indicate the
$1,\,2,\,3\sigma$ contours based on comparison with the CMB (WMAP, ACBAR, CBI)
and type 1a supernovae (Hi-Z, SCP).} 
\end{figure*}


Our search of the parameter space for the remaining models is based on a
Bayesian approach, using a Monte-Carlo Markov-Chain (\mmc) search algorithm to
identify the best cosmological models. The end product is a realization of the
posterior probability distribution function on the parameter space
\cite{Christensen:2001gj}.  Our approach is similar to the procedure described
in \cite{Verde:2003ey}, whereby the \mmc\ makes a ``smart'' walk through the
parameter space, accepting or rejecting sampled points based on a running
criteria. For each of  Q2-4, after some experimentation we found it practical
to use four independent chains in order to monitor convergence and mixing
according to the criteria of Ref. \cite{Gelman::1992}. Each such chain explored
$\sim 3\times 10^4$ models.  We found it advantageous to use a dynamical
stepsize factor $r \in[0.1,10]$. In each step, all parameters are varied
according to a Gaussian with width $r \sigma_i$, where $\sigma_i$ is an
initial  stepsize for the random walk in the direction of the $i$-th parameter.
The stepsize factor $r$ is adjusted such that the chain neither rejects nor
accepts too frequently.  Certain parameters, such as $w_0$ and $\omqls$ in the
early quintessence models, have physical boundaries and it is therefore
important that our \mmc\ does not lead to a bias in the sampled distribution
close to the boundaries. As the likelihood in the \mmc\ approach is given by
the frequency with which the chain passes through a given region of parameter
space, we would be in danger of depleting the region close to the boundary if
we were to simply discard any steps which cross the boundary.  That is, the
likelihood in the parameter region close to the boundary would be artificially
suppressed. If such a (virtual) jump in a parameter direction $i$ occurs, we
use the following strategy. We try making a gaussian-weighted step but only
towards the boundary. If this fails after ten attempts, then we start over
again by making a gaussian step in any direction. By trying ten times, however,
it is very likely  that we do make the step successfully, if we are at least
one $\sigma_i$ away from the boundary. If, however, we  are very close to the
boundary already and our step would have carried us deep into the unphysical
region, then it is sensible to make the step away from the current point, for
the chain didn't intend to stay close to the boundary anyhow. To give us
confidence in the reliability of this algorithm, we have checked by synthetic
distributions that this strategy is very well suited for sampling the
likelihood, even in cases where a physical boundary cuts off the parameter
space.


\vspace{0.2cm} \noindent{\it Q2:} We have examined quintessence models with an
equation-of-state that evolves monotonically with the scale factor, as $w(a) =
w_0 + (1-a) w_1$. For this case, the parameters are simply $\theta_Q =
\{w_0,\,w_1\}$.  This parametrization has been shown to be versatile in
describing the late-time quintessence evolution for a wide class of scalar
field models \cite{Linder:2002et}. Based on the degeneracy of models found for
Q1, we expect to find a two-dimensional family of equivalent best-fit models
with the same apparent angular size of the last scattering horizon, occupying a
plane in the $\{w_0,\,w_1,\,h\}$ space. There are three ways in which this
plane is pared down: Firstly we confine $w \ge -1$ at all times. Secondly, the
transition from $w_0$ to $w_0 + w_1$ takes place at low redshift, $z \lesssim
1$, so that high redshift supernovae restrict $w_0,\ w_1$ for these models.
Thirdly, this parametrization allows for models in which the dark energy is
non-negligible at early times, which influences the small-scale fluctuation
spectrum. The first two considerations yield $w_0 < -0.75$ at the $2\sigma$
level, marginalizing over the suppressed five dimensional parameter space, as
illustrated in Figure~\ref{fig::monotonic1}. There, the shapes of the contours
indicate that current data can only distinguish between fast ($dw/da \gtrsim
0.5$) and slow evolution of $w(a)$, and offer only a weak bound on $w_1$.
However, in terms of $\omqls$, our third consideration gives a tight upper
bound on the quintessence density during recombination. As shown in
Figure~\ref{fig::monotonic2}, $\omqls < 0.03$ at the $2\sigma$ level.  Three
target models, with significantly different equation-of-state evolution
$dw/da$, are given in Table~\ref{q2models} for future investigations.

\begin{table}[h] 
\begin{ruledtabular}
  \begin{tabular}{cccccc}
   model & $w_0$ & $w_1$ & $h$ & $\omqls$ & $\sigma_8$ \\ \hline
   Q2.1 & -0.93 & 0.43 & 0.66 & $4\times 10^{-6}$  & 0.77\\
   Q2.2 & -0.99 & 0.68 & 0.64 & $1\times 10^{-4}$  & 0.78\\
   Q2.3 & -0.92 & 0.62 & 0.62 & $1\times 10^{-4}$ & 0.73 \\
  \end{tabular}
\end{ruledtabular}
\caption{\label{q2models} Sample best-fit models with a monotonically-evolving 
equation-of-state (Q2).  Although a range of parameters give equivalently good
fits to the observational data, we have selected this sample with  $\{\Omega_b
h^2,\ \Omega_{cdm} h^2,\ n_s,\ \tau_{r}\}$ =  $\{0.023,\ 0.11,\ 0.98,\ 0.16\}$
(Q2.1), $\{0.023,\ 0.11,\ 0.98,\ 0.15\}$ (Q2.2), $\{0.023,\ 0.12,\ 0.98,\
0.08\}$ (Q2.3),  Entries for $\omqls$ and $\sigma_8$ are the resulting values
based on the other parameters. }
\end{table}


\vspace{0.2cm} \noindent{\it Q3:} We have examined models of leaping kinetic
quintessence, a scalar field evolving under an exponential potential with a
non-canonical kinetic term that undergoes a sharp transition at late times,
leading to the current accelerated expansion \cite{Hebecker:2000zb}. At early
times the field closely tracks the cosmological background with $w=0$ during
matter domination, appearing as early quintessence \cite{Caldwell:2003vp}
before undergoing a steep transition towards a strongly negative
equation-of-state by the present day. The steepness of the transition in $w$
for a leaping kinetic model is directly connected to the equation-of-state
$w_0$ today. Such models can therefore be characterized by the parameters
$\theta_{Q}=\{\omqls, w_0\}$, where $\omqls$ is the quintessence density during
recombination. (A more general parameterization, allowing for independent $w_0$
and steepness of transition can be found in \cite{Corasaniti:2002vg}.)  Since
$\omqls$ is not tied as closely to the expansion rate sampled by the
supernovae, compared to case Q2, the result is the weaker constraint $\omqls
\lesssim 0.1$, as shown in Figure~\ref{fig::leaping1}. Although the limit of a
cosmological constant can be approached in this model, the presence of early
quintessence will then require a sharp transition in the equation-of-state in
order to reach $w \to -1$. In addition to the fact that  such models lose the
early tracking behavior and instead require some degree of fine-tuning, there
is the practical consideration that the sharp transition leads to some
numerical instability in our code. To avoid this problem, we restrict $w >
-0.97$ as can be seen in Figure~\ref{fig::leaping1}. We believe that we have
succeeded at implementing an \mmc\ algorithm that does not artificially distort
the probability distribution near the parameter-space boundaries, based on our
experimentation with  synthetic distributions with boundaries. Next, because
early quintessence suppresses the growth of fluctuations on small scales
compared to large scales, we find that comparable fluctuation spectra can be
achieved by making a trade-off between $n_s$ and $\omqls$. As shown in
Figure~\ref{fig::leaping2}, slight suppression of small-scale power can be
accomplished either by a tilt towards the red, $n_s < 1$, or a rise in
$\omqls$. Since the effect of early quintessence on the small-scale fluctuation
power spectrum closely mimics a running spectral index, we have not introduced
${\rm d} n_s / {\rm d} \ln k$ as an additional parameter, which would be highly
degenerate with $\omqls$ \cite{Caldwell:2003vp}. We expect that improved
measurements of the second and third acoustic peaks will tighten the constraint
on $\omqls$ and sharpen the degeneracy in the $n_s - \omqls$ plane. Target
models, with significantly different values of early quintessence abundance,
are given in Table~\ref{q3models}.

\begin{table}[h] 
\begin{ruledtabular}
  \begin{tabular}{ccccccc}
   model & $w_0$ & $\omqls$ & $h$ & $A$ & $\bar w_{ls}$ & $\sigma_8$ \\ \hline
   Q3.1 & -0.94 & 0.006 & 0.69 & 0.0026 & -0.27 & 0.89\\ 
   Q3.2 & -0.91 & 0.024 & 0.70 & -0.0028 & -0.21 & 0.81 \\ 
   Q3.3 & -0.93 & 0.043 & 0.71 & -0.0070 &  -0.19  & 0.85 \\
  \end{tabular}
\end{ruledtabular}
\caption{\label{q3models} Sample best-fit leaping-kinetic quintessence  models
(Q3). The parameters A is an equivalent description of $\omqls$, which along
with the averaged $w$ at last scattering, $\bar w_{ls}$, can be used more
easily with equations 2-4 of  Ref.~\cite{Caldwell:2003vp} to generate the
time-evolution of these models. The non-quintessence parameters are 
$\{\Omega_b h^2,\ \Omega_{cdm} h^2,\ n_s,\ \tau_{r}\}$ = $\{0.022,\ 0.112,\
0.96,\ 0.12\}$ (Q3.1), $\{0.023,\ 0.116,\ 1.0,\ 0.16\}$ (Q3.2),  $\{0.024,\
0.119,\ 1.04,\ 0.26\}$ (Q3.3).  }
\end{table}


\vspace{0.2cm} \noindent{\it Q4:} Finally, we have examined tracker models of
quintessence. Inverse-power law (IPL) models are the archetype quintessence
models with tracking property and acceleration \cite{Ratra:1987rm,
Zlatev:1998tr,Steinhardt:nw}.  The potential is given by $V \propto
\varphi^{-\alpha}$, where the constant of proportionality is determined by
$\Omega_Q$. In certain supersymmetric QCD realizations of the IPL
\cite{Masiero:1999sq}, $\alpha$ is related to the numbers of color and flavors,
and can take on a continuous range of values $\alpha >0$. For $\alpha \to 0$,
however, inverse-power law models behave more and more like a cosmological
constant. Using earlier data, $\alpha$ has been constrained to $\alpha < 1.4$
\cite{Balbi:2000kj,Baccigalupi:2001aa}, although $h=0.65$ has been fixed in
those analyses. Keeping $h$ free, a more conservative value of $\alpha < 2$
\cite{Doran:2001ty} was inferred. From our analysis, we see that the $2\sigma$
bound has not changed dramatically. The $2\sigma$-bound with  $\alpha \lesssim
1 - 2$ is consistent with values of $h$ within the range determined by the HST,
as seen in Figure~\ref{fig::ipl1}. In   Figure~\ref{fig::ipl2} we plot the
likelihood contours in the $\Omega_m h^2 - \alpha$ plane: our results agree
with the best fit at $\Omega_m h^2 = 0.149$ for $\alpha=0$ or $w=-1$, but show
a tolerance for a wider range for $0 \le \alpha \le 2$. That is, the additional
degree of freedom in $\alpha$ means that the matter density for the IPL model
is not as well-determined from the peak position \cite{Page:2003fa} as compared
to the $\Lambda$ model. However, to maintain the peak at $\ell = 220$, we
observe that $\Omega_m h^2$ decreases slightly as $\alpha$ increases. We might
have inferred the results for the IPL based on the constant equation-of-state
models: pairs of $\{\alpha,\ h\}$ can equivalently determine a family of models
with degenerate CMB anisotropy patterns, since the differences in the late-ISW
for this model compared to Q1 make only a small contribution to the overall
$\chi^2$. Furthermore, since IPL quintessence can be modeled by the appropriate
choice of the Q2 parameters, then improved sensitivity to $dw/da$ is required
to tighten the constraints here.  While $\alpha$ is tightly constrained,  our
take-away is that IPL models with  $0.25 \lesssim \Omega_m \lesssim 0.4$ remain
viable. Target models are given in Table~\ref{q4models}.

\begin{table}[h] 
\begin{ruledtabular}
  \begin{tabular}{ccccc}
   model & $\alpha$ & $h$ & $\sigma_8$ \\ \hline
   Q4.1 & 0.1 & 0.68 & 0.90\\
   Q4.2 & 0.2 & 0.68 & 0.85\\
   Q4.3 & 0.8 & 0.68 & 0.82\\  
  \end{tabular}
\end{ruledtabular}
\caption{\label{q4models} Sample best-fit IPL quintessence  models (Q4). The
non-quintessence parameters are  $\{\Omega_b h^2,\ \Omega_{cdm} h^2,\ n_s,\
\tau_{r}\}$ = $\{0.023,\ 0.122,\ 0.97,\ 0.13\}$ (Q4.1), $\{0.023,\ 0.116,\
0.97,\ 0.14\}$ (Q4.2), $\{0.024,\ 0.102,\ 1.0,\ 0.23\}$ (Q4.3). } 
\end{table}


This work provides a capsule summary of the viable quintessence dark energy
models, based on two of the tightest constraint methods, using the CMB and SNe.
Our study advances beyond past investigations
\cite{Baccigalupi:2001aa,Brax:2000yb,Corasaniti:2001mf,Bean:2001xy,Hannestad:2002ur,Bassett:2002fe,Jimenez:2003iv,Barreiro:2003ua}
by treating a wide class of quintessence models with the powerful weight of the
WMAP data. We have considered four classes of models which cover the most basic
quintessence scenarios, including a versatile parametrization, as well as the
best motivated and most realistic scenarios based on our current understanding
of particle physics. Absent from our survey are {\it k}~essence models, and
dark energy models with a coupling to other matter fields or gravity. In the
former case, since the sound speed of fluctuations varies with time in these
models, however, we can make a simple distinction with quintessence models with
an underlying scalar field, wherein the propagation speed is equal to the speed
of light. (See Refs. \cite{Erickson:2001bq,DeDeo:2003te} for analysis of these
models with respect to CMB anisotropy.) For the latter case we refer to Ref.
\cite{Amendola:2003eq} for specific coupled models.  We also note that the mass
fluctuation power spectrum is an important cosmological constraint which we
have omitted at this stage, primarily because it constrains energy density
rather than pressure (although there are exceptions), a chief feature
distinguishing dark energy from dark matter. (The constraint curves obtained in
Ref. \cite{Schuecker:2002yj}, {\it e.g.} Figure 3, are consistent with, but do
not decisively pare down the parameter regions of our current results.) 
Furthermore, the analysis of mass power spectrum observations will have to take
into account possible influences of a time-dependent $w$ and early
quintessence, which we put off for later investigation. Overall, we have
simulated in detail more than 400,000 individual cosmological models. The
stored spectra and parameter-space likelihood functions will be used to
evaluate additional constraints that can offer clues to the behavior of the
dark energy.  The set of thirteen target models, listed in
Tables~\ref{q1models}-\ref{q4models},  which includes a revised concordance
$\Lambda$ model \cite{Wang:1999fa}, should be of use by researchers to further
test the quintessence hypothesis.


\noindent {\bf Acknowledgments}\, This work was supported by NSF grant
PHY-0099543 at Dartmouth. We thank Pier Stefano Corasaniti for useful
conversations, and Dartmouth colleagues Barrett Rogers and
Brian Chaboyer for use of computing resources.



\end{document}